\documentclass{jltp}

\usepackage{graphicx} 

\title{Adiabatic Phase Diagram on Degenerate Fermi Gas 
with Feshbach-Resonance}

\author{S. Watabe, T. Nikuni$^{*}$, 
N. Nygaard$^{\dag}$, J. E. Williams$^{\ddag}$, and C. W. Clark$^{\ddag}$}

\address{Department of Physics, Graduate School of Science, University of Tokyo,\\ Tokyo, Japan\\$^*$Department of Physics, Faculty of Science, Tokyo University of Science,\\ Tokyo, Japan\\$^{\dag}$Danish National Research Foundation Center for Quantum Optics,\\ Department of Physics and Astronomy, University of Aarhus,\\ DK-800 {\AA}rhus C, Denmark\\$^{\ddag}$
Electron and Optical Physics Division,\\
National Institute of Standards 
and Technology, Gaithersburg,\\ MD 20899-8410, USA
}

\runninghead{S. Watabe {\it et al.}}{Adiabatic Phase Diagram}

\begin{document}

\maketitle

\begin{abstract}
We determine the adiabatic phase diagrams 
for a resonantly-coupled system of Fermi atoms 
and Bose molecules confined in a harmonic trap 
by using the local density approximation. 
The key idea of our work is conservation of entropy 
through the adiabatic process. 
We also calculate the molecular conversion efficiency 
as a function of the initial temperature. 
Our work helps to understand 
recent experiments on the BCS-BEC crossover, 
in terms of the initial temperature 
measured before a sweep of the magnetic field.

\end{abstract}

\section{Introduction}
Degenerate Fermi gases with controllable interaction 
have been realized in resent experiments 
by making use of a Feshbach resonance. 
By ramping a magnetic field across the resonance, 
one can observe crossover between 
BCS and BEC superfluidity in a Fermi gas.\cite{regal,zwierlein} 
Williams {\it et al}.\cite{williams} calculated equilibrium phase diagrams 
for an ideal gas mixture of fermionic atoms 
and bosonic molecules, 
which provide qualitative understanding of experimental data 
of Refs.\onlinecite{regal,zwierlein}. 
In this paper, 
we extend the work by 
Williams {\it et al}.\cite{williams} 
to include the resonant interaction 
using mean-field theory. 
Calculating the entropy as a function of the temperature 
and the resonance energy, 
we show the paths of constant entropy 
traversed in the conventional phase diagrams 
as the resonance energy is varied adiabatically. 
On the basis of the adiabatic path, 
we determine the adiabatic phase diagrams 
against the resonance energy and entropy, 
labeled by the initial temperature of the Fermi
gas measured before a sweep of the magnetic field. 
We also calculate 
the production efficiency of molecules 
during the adiabatic sweep of the magnetic field, 
by extending the calculation given in 
the recent work by Williams {\it et al}.\cite{williams-conversion} 
to include mean-field interactions. 
Our study gives an intuitive understanding 
of recent experiments on the BCS-BEC crossover 
using the adiabatic sweep process. 

\section{Equilibrium Theory and Phase Diagrams}
The starting point of our theory is the coupled boson-fermion  Hamiltonian
 with two-component Fermi gas.\cite{falco,matt,kawaguchi} 
Our mean-field calculation 
closely follows the procedure described in detail in Ref.\onlinecite{matt}. 
The resulting grand canonical potential $\Omega$ 
is given by 
\begin{eqnarray}
\Omega &=& 
\int d{\bf r} 
\left  (
[\varepsilon_{{\rm res}}-2\mu({\bf r})]|\Phi({\bf r})|^{2}
+ 
\frac{1}{\beta}
\int \frac{d{\bf k}}{(2\pi^{3})}
\ln{ \{1-e^{-\beta[\varepsilon_{\bf k}^{(m)}-\varepsilon_{{\rm res}}-2\mu({\bf r})]}\} }
\right .
\nonumber
\\
 & & 
+
\left .
\int \frac{d{\bf k}}{(2\pi^{3})}
\left \{
[\varepsilon_{\bf k}^{(a)}-\mu({\bf r})-E_{\bf k}({\bf r})]
-
\frac{2}{\beta}
\ln{[1+e^{-\beta E_{\bf k}({\bf r})}]}
\right \}
\right ) .
\end{eqnarray}
The chemical potential $\mu$ is introduced to impose the constraint that 
the number of particles be conserved. 
The effect of a harmonic trap 
is included by the local density approximation,\cite{ohashi} 
which adopts the effect of local potential 
by replacing the chemical potential $\mu$ 
with $\mu({\bf r}) = \mu - V_{a}({\bf r})$, 
where $V_{a}({\bf r})$ is the harmonic trap for fermionic atoms. 
$ \varepsilon^{\scriptsize{(m)}}
_{{\bf k}}$
is the kinetic energy of a molecule, 
and 
$ \varepsilon^{\scriptsize{(a)}}
_{{\bf k}}$
is the kinetic energy of an atom. 
The local quasiparticle excitation energy is
$E_{{\bf k}} ({\bf r})= 
\sqrt{[\varepsilon_{{\bf k}}
^{(a)}-\mu ({\bf r})]^{2}
+|\Delta ({\bf r})|^{2}}$, where the local gap energy is
 $|\Delta ({\bf r})| = \kappa |\Phi ({\bf r})|$, 
$\Phi({\bf r})$ is the condensed molecular wavefunction, 
and $\kappa$ is the coupling constant of the resonance interaction. 
Finally, 
$\varepsilon_{{\rm res}}$ is the bare resonance energy, 
which is controlled by the magnetic field. 

We regard the equation of the number of particles and the gap equation 
as the simultaneous equations for 
the chemical potential $\mu$ and the local gap $|\Delta ({\bf r})|$, 
for a given temperature $T$ and a resonance energy  $\varepsilon_{{\rm res}}$. 
In this paper, 
we consider the case of narrow resonance. 
We set the bare coupling strength to 
$\alpha \equiv \sqrt{\rho} \kappa = 0.4 \varepsilon_{\rm{F}}$, 
where $\rho$ is the peak density and $\varepsilon_{\rm{F}}$ is the Fermi energy 
of a pure gas of fermionic atoms. 
To avoid the divergence in the gap equation, 
we impose the gaussian cut-off. 

\begin{figure}[htbp]
\begin{minipage}{0.5\hsize}
\begin{center}
\includegraphics[width=2in,height=2in,keepaspectratio,clip]{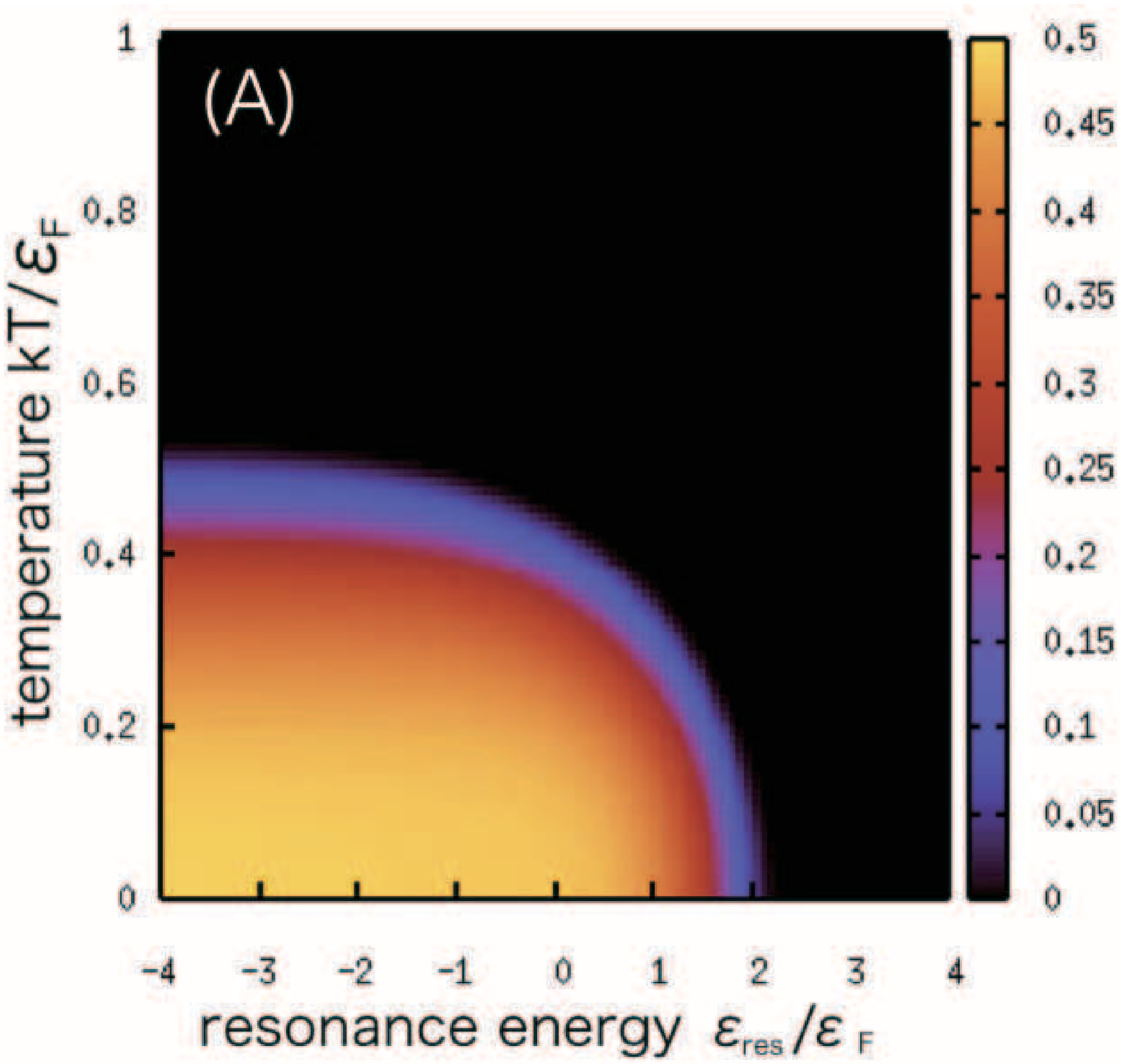}
\end{center}
\end{minipage}
\begin{minipage}{0.5\hsize}
\begin{center}
\includegraphics[width=2in,height=2in,keepaspectratio,clip]{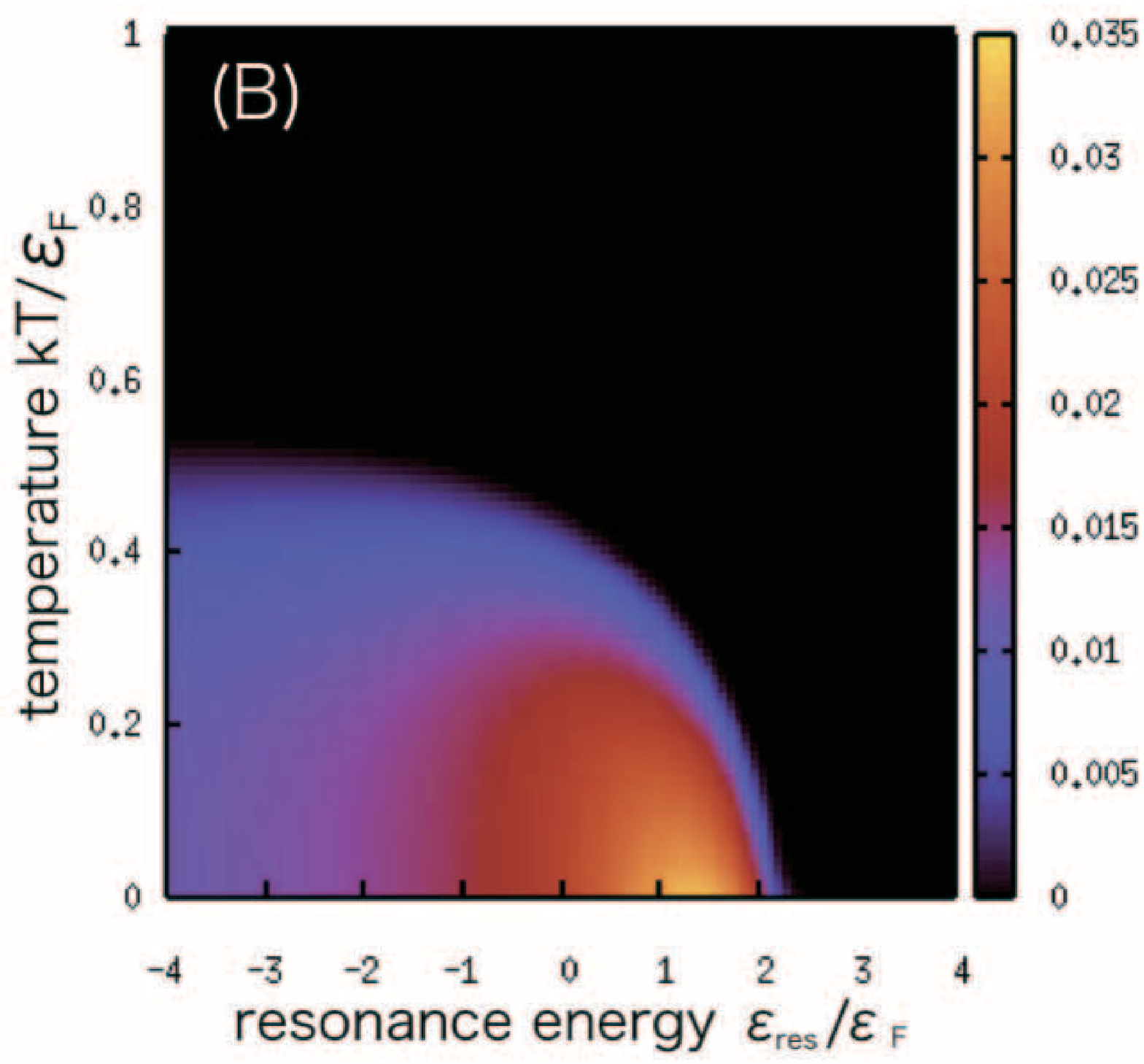}
\end{center}
\end{minipage}
\caption{Phase diagrams of 
the condensed molecular fraction $\eta_{mc}$ (A)
and 
the Cooper pair fraction $\eta_{p}$ (B). }
\label{LDATNmcNp.fig}
\end{figure}
The condensed pair number $N_{c}$ is composed of the number of 
condensed molecules $N_{mc}$ and Cooper pairs $N_{p}$. 
In Fig.\ref{LDATNmcNp.fig} (A), 
we plot the condensed molecular fraction $\eta_{mc} = N_{mc}/N_{\rm tot}$ 
against the temperature $T$ and 
the resonance energy $\varepsilon_{{\rm res}}$. 
We also plot the condensed Cooper pair fraction 
$\eta_{p} = N_{p}/{N_{\rm tot}}$, 
which was defined by Yang\cite{yang}, 
in Fig.\ref{LDATNmcNp.fig} (B). 
The energy scale is normalized 
by the Fermi energy $\varepsilon_{\rm{F}}$ 
of a pure atomic Fermi gas. 
We do not present 
the phase diagram for the molecular fraction 
$\eta_{m} = N_{m}/{N_{\rm tot}}$, 
since it is very similar to that 
given by Williams {\it et al}.\cite{williams} for an ideal gas. 

\section{Adiabatic Phase Diagrams} 
In this section, we discuss the adiabatic phase diagrams. 
We first briefly review the experimental procedures.\cite{regal,zwierlein} 
In these experiments, the ultracold two-component Fermi gas 
is initially prepared at a magnetic field 
detuned far from the resonance position, corresponding to a pure atomic gas. 
In this weakly interacting regime, the temperature of Fermi gas is measured 
using time-of-flight imaging.
The magnetic field is then slowly lowered to the vicinity of the resonance
 to allow the atoms and molecules sufficient time to move and collide 
in the trap. 
This process may satisfy the condition 
that the gas be able to collisionally relax to equilibrium.\cite{williams} 
We regard this process as an adiabatic quasistatic process.
The system will follow a path of constant entropy in the phase diagram. 

\begin{figure}[htbp]
\begin{minipage}{0.5\hsize}
\begin{center}
\includegraphics[width=2in,height=2in,keepaspectratio,clip]{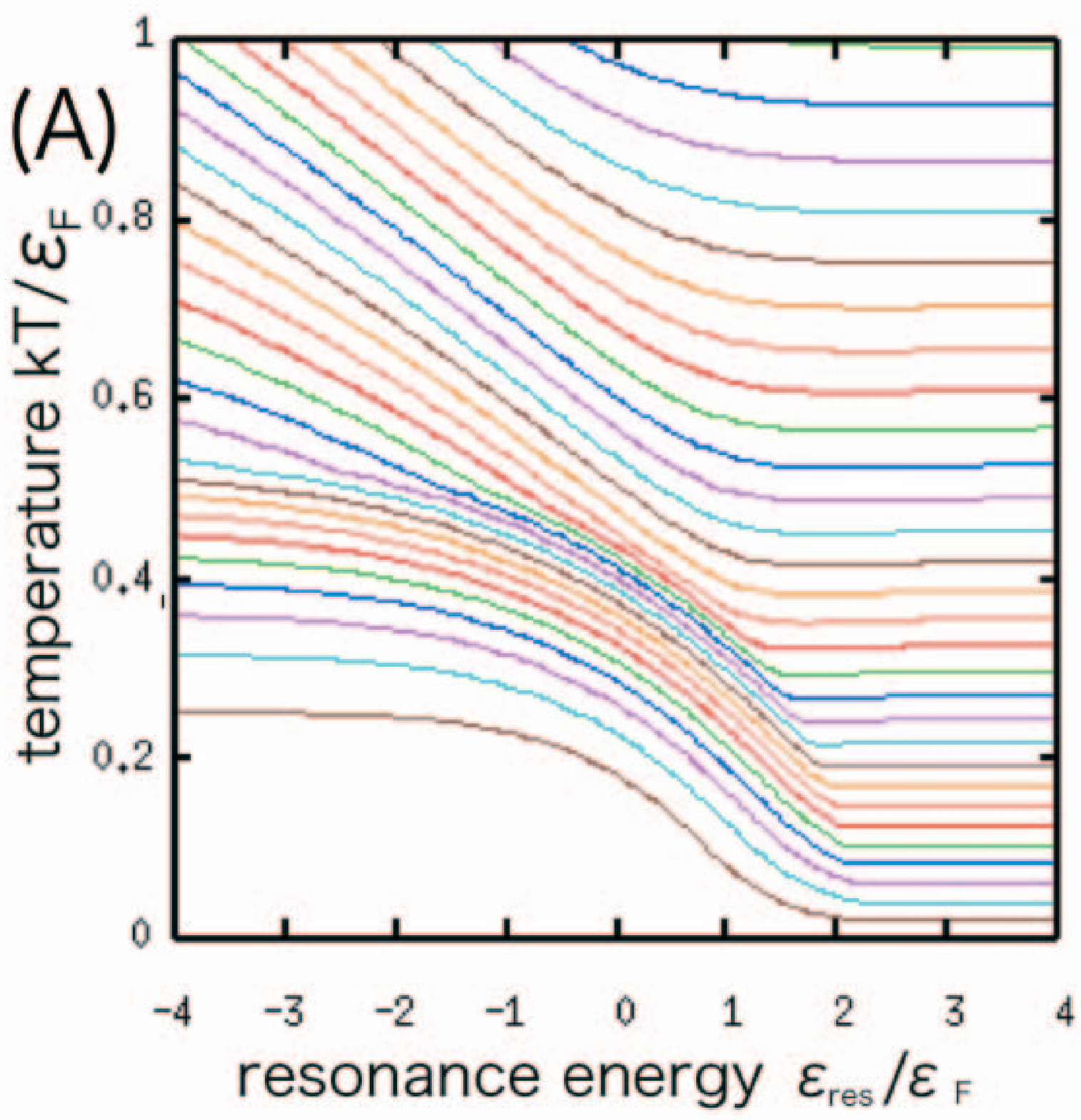}
\end{center}
\end{minipage}
\begin{minipage}{0.5\hsize}
\begin{center}
\includegraphics[width=2in,height=2in,keepaspectratio,clip]{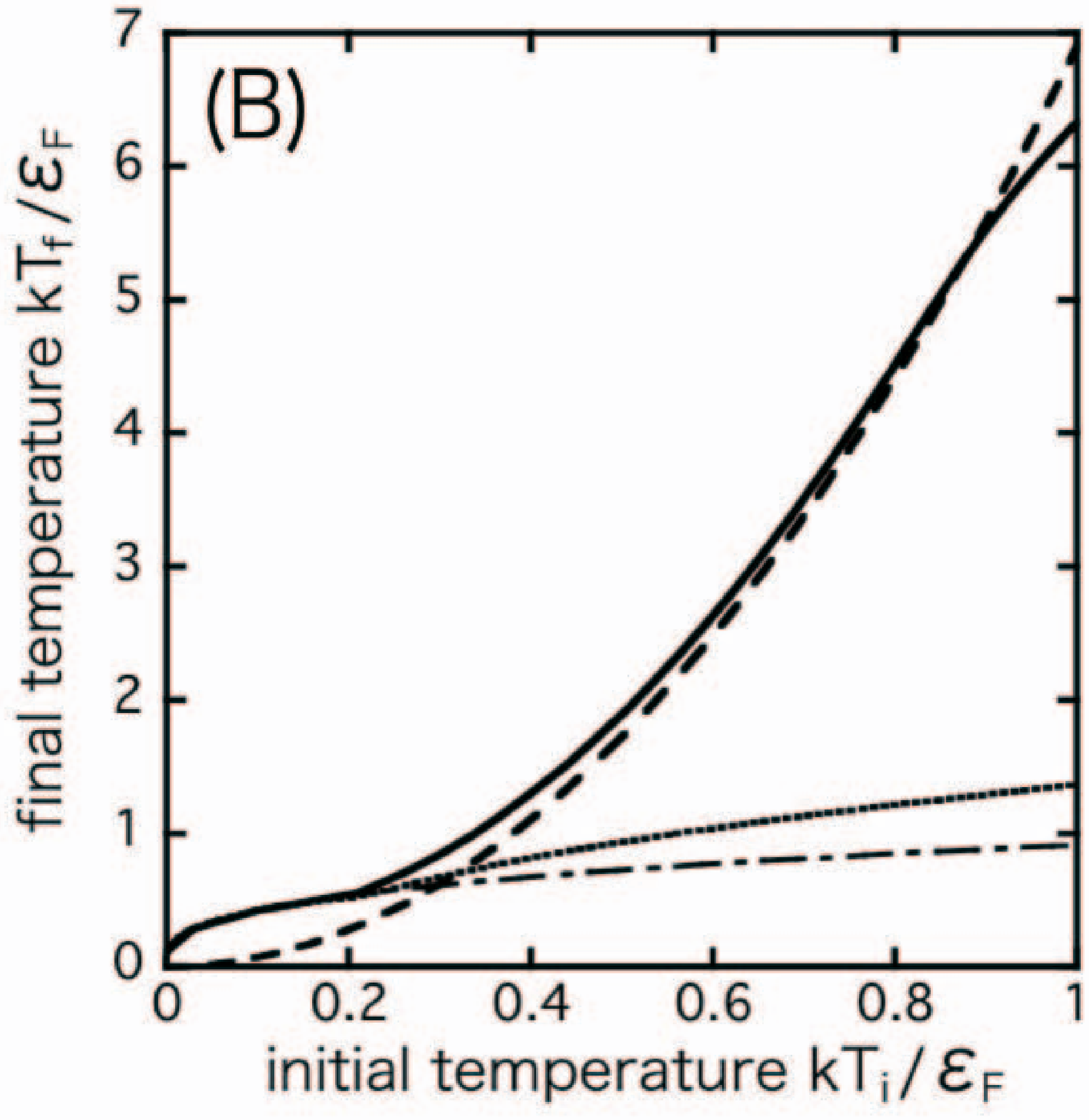}
\end{center}
\end{minipage}
\caption{Contours of constant entropy (A) and 
final temperature versus initial temperature (B).}
\label{LDA2.fig}
\end{figure}

In Fig.\ref{LDA2.fig}(A), we show contours of 
constant entropy, which is the sum of the atom entropy and 
the molecule entropy, 
in the $\varepsilon_{{\rm res}}$ - $T$ plane. 
We recognize those lines as the paths 
that are traversed 
as the resonance energy $\varepsilon_{{\rm res}}$ 
is swept adiabatically. 

The relation between the initial temperature 
of the atomic gas 
and the final temperature of the molecular gas 
has been given for 
the ideal gas mixture of fermionic atoms and 
bosonic molecules by Williams {\it et al}.\cite{williams} 
This relation can be derived 
by connecting the initial entropy 
of the atomic gas 
and the final entropy of the molecular gas. 

In Fig.\ref{LDA2.fig}(B), we plot the final temperature 
in a deeply BEC region and a moderately BEC region 
as a function of the initial temperature. 
The solid line is obtained by a numerical calculation 
in a deeply BEC region 
where $\varepsilon_{{\rm res}} = -100\varepsilon_{\rm{F}}$. 
The dotted line is obtained by a numerical calculation 
in a moderately BEC region 
where $\varepsilon_{{\rm res}} = -4\varepsilon_{\rm{F}}$. 
The dashed line represents the high-temperature limit 
for an ideal gas, 
while the dot-dashed line represents the low-temperature limit 
for an ideal gas. 
The low-temperature approximation 
agrees with the both of numerical solutions. 
The high temperature approximation does not agree with 
the numerical solution for a moderately BEC region, 
because atoms appear 
when the temperature is not sufficiently high 
and thus the concept of connecting 
the initial entropy of atoms and 
the final entropy of molecules 
breaks down. 
On the other hand, 
the high temperature approximation 
agrees with a numerical solution 
of a deeply BEC region 
except for the high initial temperature regime 
where atoms appear. 

\begin{figure}[htbp]
\begin{minipage}{0.5\hsize}
\begin{center}
\includegraphics[width=2in,height=2in,keepaspectratio,clip]{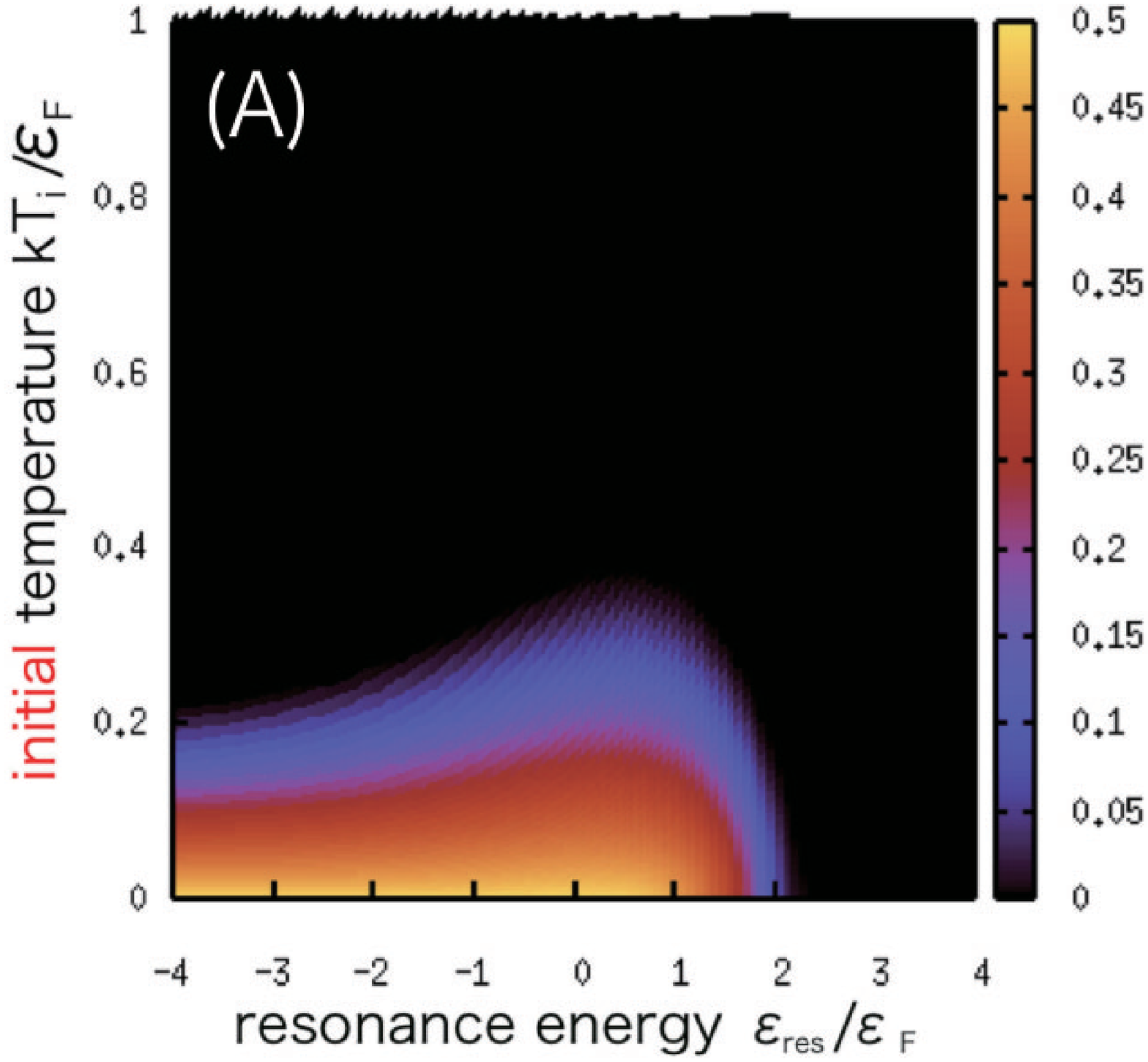}
\end{center}
\end{minipage}
\begin{minipage}{0.5\hsize}
\begin{center}
\includegraphics[width=2in,height=2in,keepaspectratio,clip]{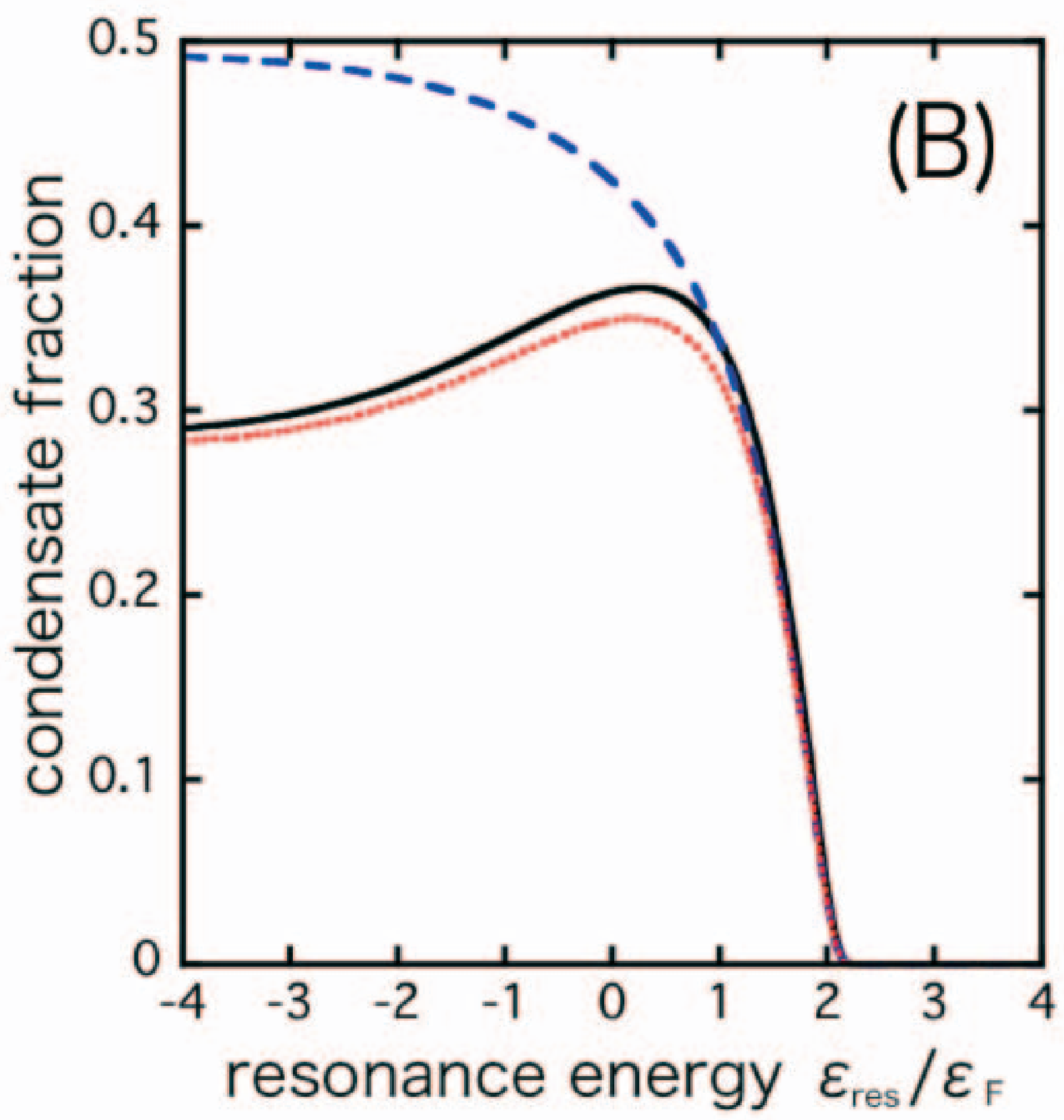}
\end{center}
\end{minipage}
\caption{Adiabatic phase diagram (A) and 
each fraction 
for initial temperature $T_{\rm i} = 0.08T_{\rm F}$ (B). }
\label{LDA3.fig}
\end{figure}

To compare our results 
with the phase diagrams of the experiments
\cite{regal,zwierlein} directly, 
we plot the fermionic condensate fraction $\eta_{c}$ 
against the resonance energy $\varepsilon_{{\rm res}}$ 
and the initial temperature of a gas 
in Fig.\ref{LDA3.fig}(A). 
The initial temperature $T_{i}$ is found 
by equating the total entropy to the initial entropy 
$S_{\rm tot}(\varepsilon_{{\rm res}}, T) = S_{i}(T_{i})$. 
In Fig.\ref{LDATNmcNp.fig}, 
the transition temperature is 
$T_{c} \simeq 0.53T_{F}$ in the BEC region 
where $\varepsilon_{{\rm res}} = -4 \varepsilon_{\rm{F}}$. 
On the other hand, in Fig.\ref{LDA3.fig}(A), 
the transition temperature in terms of 
the initial temerature is 
$T_{i,c} \simeq 0.21T_{F}$ at $\varepsilon_{{\rm res}} = -4 \varepsilon_{\rm{F}}$. 
This translated transition temperature $T_{i,c}$ 
in the BEC limit 
is comparable with the transition temperature observed in the resent experiments.
\cite{regal,zwierlein} 

In Fig.\ref{LDA3.fig}(B), 
we plot 
the fermionic condensate fraction (black solid line) 
and the condensate molecular fraction (red dotted line) 
with an initial temperature $T_{i}/T_{F} = 0.08$ 
as a function of the resonance energy. 
The blue dashed line represents the molecular fraction. 
The behavior of the condensate fraction shown in Fig.\ref{LDA3.fig}(B) 
agrees with the experiments.\cite{regal,zwierlein} 

\section{Molecular Conversion Efficiency}
Hodby {\it et al}.\cite{hodby} 
have investigated 
the molecular conversion efficiency. 
They found that the molecular conversion efficiency 
does not reach 100\% even for an adiabatic sweep, 
but saturates at a value that 
depends on the initial peak phase-space density. 

\begin{figure}
\begin{center}
\includegraphics[width=2in,height=2in,keepaspectratio,clip]{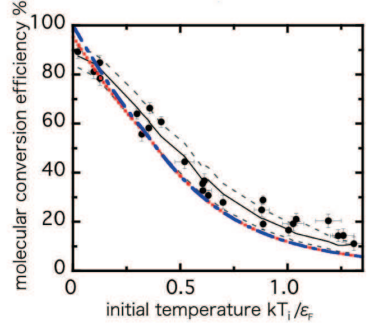}
\end{center}
\caption{The molecular conversion efficiencies.
}
\label{LDAMCE.fig}
\end{figure}

Assuming that the principle controlling the maximum molecule conversion efficiency proposed by 
Williams {\it et al}.\cite{williams-conversion} 
also applies when including 
the resonant interaction, 
we calculate the molecular conversion efficiency 
using $\eta_{m}(\varepsilon_{{\rm res}} = 0)$ 
as a function of initial temperature, 
as shown in Fig.\ref{LDAMCE.fig}. 
The red dotted line represents the result from our model, 
while the blue dashed line represents the ideal gas mixture 
of Fermi atoms and Bose molecules. 
The black dots represent 
the data of Hodby  {\it et al}.,\cite{hodby}  
and the black solid line represents their simulation result. 
Black dotted lines represent the uncertainty 
of the pairing parameter 
in their simulation. 
Our model agrees well with the experimental data.

The resonant interaction is more important 
for lower initial temperatures. 
At the initial temperature $T_{i} = 0$, 
the molecular conversion efficiency is 100\% 
in the ideal gas mixture model. 
On the other hand, 
the experimental data 
in the low temperature limit $T_{i}\rightarrow 0$ 
is suppressed from 100\%. 
The efficiency in our model at $T_{i}=0$ 
is also less than 100\%, 
because the resonant interaction suppresses the molecular conversion. 
\section{Conclusion}
The adiabatic phase diagram 
allows us to compare the theory with the experiments, 
since in the experients,\cite{regal,zwierlein} 
the bimodal condensate fraction 
have been measured against the initial temperature and the magnetic field. 
We obtain the transition temperature 
in terms of the initial temperature 
$T_{i,c} \simeq 0.21T_{F}$ in the BEC regime, 
which is consistent with the experimental result.\cite{regal,zwierlein} 
Finally, the molecular conversion efficiency 
is plotted as a function of the initial temperature. 
We find that the resonant interaction suppresses 
the complete conversion.

\end{document}